\documentclass[usenatbib,usegraphicx]{mn2e}
\usepackage{natbib}
\bibpunct{(}{)}{;}{a}{}{,}
\usepackage{graphicx}
\usepackage{psfig}
\usepackage{times}
\usepackage{amsmath}
\usepackage{amssymb}

\title[Oscillating relativistic tori around Kerr black holes]
{Dynamics of oscillating relativistic tori around  Kerr
  black holes}

\author[Zanotti et~ al.]
	{Olindo Zanotti$^{1}$, Jos\'e A. Font$^{1}$,        
	Luciano Rezzolla$^{2,3}$, Pedro J. Montero$^{2}$  \\
								\\
	$^{(1)}$Departamento de Astronom\'{\i}a y Astrof\'{\i}sica,
	Universidad de Valencia, Dr. Moliner 50, 46100 Burjassot
	(Valencia), Spain                               	\\ 
        $^{(2)}$SISSA, International School for Advanced Studies and INFN,
        Via Beirut, 2 34014 Trieste, Italy              \\
        $^{(3)}$Department of Physics and Astronomy, Louisiana State
        University, Baton Rouge, LA 70803 USA           \\
	}

\pagerange{\pageref{firstpage}--\pageref{lastpage}}
\pubyear{2004}

\begin{document}

\maketitle
\label{firstpage}

\begin{abstract}
	We present a comprehensive numerical study of the dynamics of relativistic
	axisymmetric accretion tori with a  power-law distribution of specific  
        angular momentum orbiting in the background spacetime of a Kerr black  
        hole.  By  combining  general   relativistic  hydrodynamics
	simulations  with   a  linear  perturbative
	approach  
	we investigate the main
	dynamical  properties of  these  objects over  a large  parameter
	space. The  astrophysical implications of our  results extend and
	improve  two   interesting  results  that   have  been  recently reported
        in the literature.  Firstly,  the induced quasi-periodic  variation of the
	mass quadrupole  moment  makes  relativistic  tori of  nuclear  matter
	densities, as those formed during the last stages of binary neutron star 
        mergers, promising  sources of gravitational radiation, potentially
	detectable  by interferometric  instruments.   Secondly, $p$-mode
	oscillations in relativistic tori of low
	rest-mass densities could
	be  used to  explain high  frequency  quasi-periodic oscillations
	observed in X-ray binaries containing a black hole candidate
	under conditions more generic than those considered so far.

\end{abstract}


\date{Accepted 0000 00 00.
      Received 0000 00 00.}

\section{Introduction}
\label{intro}

	Relativistic tori (i.e. geometrically thick discs) orbiting
around black holes are expected to form in a number of different
scenarios, such as after the gravitational collapse of the core of a
rotating massive star ($M>25 M_{\odot}$) or after a neutron star binary
merger. State-of-the-art numerical simulations of these scenarios, both
in Newtonian physics \citep{lee:01,ruffert:01}, as well as
in a relativistic framework \citep{shibata:03}, have shown that 
under certain conditions, a massive disc may be produced. This disc 
surrounds the newly formed black hole, has large average rest-mass 
densities and highly super-Eddington mass fluxes, and its angular 
momentum obeys a sub-Keplerian distribution. Accretion tori of much 
smaller accretion rates have also been considered in general relativistic 
magnetohydrodynamic simulations of accretion flows onto Kerr black 
holes \citep{devel:03,harm,gammie:04}.

	There are several reasons for considering these objects
astrophysically interesting. One of the most important reason has to do
with the idea that an accreting torus around a black hole generated after
neutron star coalescence could be the progenitor source for short
duration, (i.e. $<2$ s) gamma-ray bursts (GRBs), where the relativistic
fireball expands along the rotation axis with very low baryonic
contamination~\citep{goodman:86,paczynski:86} (see \citet{aloy:04} for 
recent relativistic hydrodynamic simulations of such a scenario). 
Correspondingly, in the ``collapsar'' scenario, the core of a massive rotating 
star collapses to form a rotating black hole~\citep{woosley:01} which subsequently grows accreting the surrounding envelope of the star
through a geometrically thick disc.
An energetic,
long-duration (i.e. $>2$ s) GRB accompanied by a Type Ib/Ic supernova may
be produced as a result of such (thick) disc accretion process. As in the
neutron star merging scenario, part of the large amounts of the energy
released by accretion is deposited in the low-density region along the
rotation axis of the star, where the heated gas expands in a jet-like
fireball (see \citet{macfadyen:99,aloy:00} for numerical simulations of
this process).

	Even in the absence of other potential instabilities triggered
by nonaxisymmetric perturbations or magnetic fields, the accretion torus that
forms in the aforementioned catastrophic events is likely to be accreting on to
the black hole with rates which largely exceed the Eddington rate, a property 
which may lead to the rapid destruction of the disc. Such a possibility was 
first suggested by \citet{abramowicz:83}, who pointed out how the equilibrium 
solution of a barotropic fluid orbiting a black hole with a general, non-Keplerian 
rotation law might be hydrodynamically unstable. The outermost closed equipotential 
surface of the torus has a distinctive cusp at the inner edge through which mass
transfer can be driven by small changes in the pressure gradient there (in a 
stationary situation the pressure gradient at the cusp would be exactly zero). 
The instability is driven by the progressive penetration
of the cusp into the disc as a result of the accretion of mass and
angular momentum and of the corresponding increase of mass and spin of
the black hole. During the development of the instability the mass flux
increases exponentially and leads to the complete disappearance of the
disc into the black hole on a dynamical timescale, justifying why it
is usually referred to as the ``runaway instability''.

	Axisymmetric, general relativistic hydrodynamical simulations of 
this process have been performed only recently 
\citep{font:02,font2:02,zanotti:03,daigne:04}. In these works, the disc
is assumed to be non-selfgravitating and the relativistic hydrodynamics
equations are then evolved in time in the curved spacetime of the black
hole. The latter is instantaneously modified to new stationary
solutions as the mass and spin of the black hole gradually increase as a
result of accretion. While this is clearly an approximation to solving
the full Einstein equations, it is a reasonable one when the disc-to-hole 
mass ratio is {\it small enough} and the changes in the spacetime between 
two successive timelevels is correspondingly small. Within this approximation 
the numerical simulations have shown that while thick discs with a 
{\it constant} distribution of the specific angular momentum are subject 
to the runaway instability irrespective of the initial conditions 
\citep{font:02,zanotti:03}, a small departure from a constant distribution 
of angular momentum (namely, a power-law increase with the radial distance 
with a small index) suffices to prevent the development of the 
instability~\citep{font2:02,daigne:04}. This result is in agreement with 
perturbative calculations of radially oscillating tori which indicate
that the oscillations have eigenfunctions that are significantly modified 
at the inner edge when the specific angular momentum is not constant 
\citep{ryz:03}. However, no definitive conclusion can be made yet on the 
occurrence of the runaway instability as there are at least two other physical 
processes that could play some role. One of these is provided by the self-gravity 
of the disc which, from studies based on sequences of stationary models has been
shown to favour the instability \citep{masuda:98}; the other one is brought up
by the presence of magnetic fields that would affect globally the properties 
of accretion flows and the angular momentum transport in discs surrounding 
black holes.

	Magnetic fields are indeed expected to influence significantly
the physics of accretion discs, either geometrically thin or thick, in
many respects, well beyond the issue of the runaway instability. In
particular, the impact of magnetic fields in the dynamics of accretion
flows has been recognized in a large number of works, but only recently the
first magnetohydrodynamical simulations in the relativistic regime have become
available (but see \citet{yokosawa:95} for earlier investigations). A number 
of authors \citep{mck:02,devel:03,harm,hirose:03,gammie:04} have
started programs to perform realistic simulations of magnetically driven 
accretion flows in rotating black hole spacetimes. A fundamental
idea under intense scrutiny is that the combined presence of a magnetic
field and of a differentially rotating fluid, in particular a Keplerian
disc, can lead to the so called magnetorotational instability (see
\citet{balbus:03} for a review), generating an effective viscosity and
thus transporting angular momentum outward through magnetohydrodynamic
turbulence.

	Notwithstanding that magnetic fields play an important role in
accretion theory, two new interesting results have emerged recently in
purely hydrodynamical axisymmetric simulations.
Both of them refer to the evidence that upon the
introduction of perturbations stable relativistic tori manifest a
regular, long-term oscillating behaviour. This, in turn, is responsible
for the periodic changes of the mass quadrupole moment of the discs, thus 
making these objects promising new sources of detectable gravitational radiation.
This is particularly true when the average density in the discs is close to nuclear
matter density~\citep{zanotti:03}, as it is the case in discs formed as a
result of binary neutron star coalescence. In addition, when the thick
discs are composed of the low-density material stripped from the
secondary star in low-mass X-ray binaries (LMXBs), their oscillations
could help explaining the high frequency quasi-periodic oscillations (HFQPOs) 
observed in the spectra of X-ray binaries~\citep{ryz:03}. In a model proposed
recently and based on the evidence that tori around black holes have the
fundamental mode of oscillation and the first overtones
in the harmonic sequence $2:3:4\ldots$ to a good precision and
in a very wide parameter space, the HFQPOs are explained
in terms of $p$-mode oscillations of a small-size
accretion torus orbiting around the black hole (see also
Abramowicz \& Klu\'zniak 2004 for a recent review on the
phenomenology associated with HFQPOs and for alternative models).

	The aim of this paper is to extend the analysis of the dynamics
of axisymmetric relativistic tori carried out by \citet{zanotti:03}
(henceforth paper I), in which the attention was limited to models with
{\it constant} specific angular momentum orbiting around a {\it
Schwarzschild} (nonrotating) black hole. Now we consider the more general
case of {\it nonconstant} angular momentum thick discs in the rotating
background spacetime provided by the {\it Kerr} metric. Furthermore,
since the results reported recently by~\citet{daigne:04} pose limits on the
occurrence of the runaway instability in the case of non
self-gravitating, nonconstant angular momentum discs, we do not
investigate this issue further here. Instead, we focus on the study of the 
oscillating behaviour of marginally stable discs. To this aim we combine two
complementary tools: nonlinear hydrodynamic simulations and a linear
perturbative approach. With the first one we can follow the long-term
evolution of perturbed thick accretion discs, analyze their dynamics, and
compute the gravitational wave emission in the Newtonian quadrupole
approximation. Correspondingly, with the second approach we can
investigate how the oscillation properties of these objects depend on
their geometrical features, surveying a very large parameter space. When
applied to vertically integrated relativistic tori, this second approach
has already shown its utility for explaining the results of the numerical 
simulations of \citet{font:02,font2:02} and of \cite{zanotti:03}.

	The paper is organized as follows: In Section~\ref{II} we
briefly review the main properties of relativistic tori which obey a
power-law angular momentum distribution. Next, in Section~\ref{III}, we
present the two mathematical approaches used in our investigation, namely
the techniques for the numerical solution of the hydrodynamic equations
and the basic aspects related with the solution of the eigenvalue problem
in the linear perturbative framework.  Section~\ref{IV} presents the
initial models, while Section~\ref{V} contains the discussion on the
results. Finally, Section~\ref{VI} is devoted to the conclusions.
Throughout the paper we use a space-like signature $(-,+,+,+)$ and a
system of geometrized units in which $G = c = 1$. The unit of length is
chosen to be the gravitational radius of the black hole, $r_{\rm g}
\equiv G M/c^2$, where $M$ is the mass of the black hole. Greek indeces 
run from 0 to 3 and Latin indeces from 1 to 3. 

\section{Stationary fluid configurations}
\label{II}

	The procedure for building a stationary and axisymmetric fluid
configuration orbiting around a Kerr black hole and obeying a power-law
distribution of the specific angular momentum in the equatorial plane has
been reviewed in great detail by \citet{daigne:04}. Here, for the sake of 
completeness, we will only recall those definitions and physical properties 
that are essential to the present study. In what follows we consider a 
perfect fluid described by the stress-energy tensor
\begin{equation}
\label{stress-tensor}
T^{\mu\nu}\equiv (e+p)u^\mu u^\nu+p g^{\mu\nu} = \rho h u^\mu u^\nu+p
	g^{\mu\nu} \ ,
\end{equation}
where $g^{\mu\nu}$ are the coefficients of the Kerr metric in
Boyer-Lindquist coordinates $(t,r,\theta,\phi)$ and $e$, $p$, $\rho$, and
$h = (e+p)/\rho$ are the energy density, the isotropic pressure, the
rest-mass density, and the specific enthalpy, respectively, each of them
measured in the frame comoving with the fluid. The fluid is supposed to
obey a polytropic equation of state $p=\kappa \rho^\gamma$ (EOS), where
$\kappa$ is the polytropic constant and $\gamma$ is the adiabatic
index. We use $\Omega \equiv u^{\phi}/u^t$ to denote the angular velocity
of the fluid and $\ell \equiv - u_\phi/u_t$ to denote its specific angular
momentum. Moreover, the rotation law on the equatorial plane is
given by a power-law distribution of the specific angular momentum
\begin{equation}
\label{power_law}
\ell (r, \theta = \pi/2) = {\cal S} r^q \ ,
\end{equation}
where ${\cal S}$ can be either positive or negative, according to the disc
rotation being prograde or retrograde, respectively, with respect to the
black hole rotation. When the motion of the fluid is just circular and
does not change in time (i.e. $\partial_t = 0 = \partial_{\phi}$), the
relativistic Euler equations in the $r$ and $\theta$ directions are given
by the (Bernoulli-type) equilibrium conditions
\begin{equation}
\label{bernoulli}
\frac{\nabla_i p}{e+p} = - \nabla_i \ln(u_t) +
	\frac{\Omega \nabla_i \ell}{1- \Omega \ell} \hspace{1cm}
	(i=r,\theta) \ ,
\end{equation} 
where the right-hand-side is nothing but the opposite of the relativistic
four acceleration $a_i \equiv u^\alpha\nabla_\alpha u_i$, which vanishes
in the case of a purely geodetic motion\footnote{Note that
Eq.~(\ref{bernoulli}) implies that a stationary and axisymmetric circular
motion of a perfect fluid is geodesic if and only if it follows a
Keplerian rotation law.}. The procedure for solving Eq.~(\ref{bernoulli})
exploits the fact that the surfaces of constant $\Omega$ (the so-called
von Zeipel cylinders) coincide with the surfaces of constant $\ell$ (see
\citet{daigne:04} for details), a result which holds true only for a
barotropic fluid \citep{abramowicz:71}. One of the most relevant features
of the equilibrium solution, irrespective of the
distribution of the angular momentum, is that the resulting
thick disc can have two Keplerian points on the equatorial plane
(i.e. points where the rotation law is that of the Keplerian motion): the
first one is the cusp, through which matter can accrete onto the black
hole, and the second one marks the position of the maximum rest-mass
density.  However, while in the case of a constant angular momentum
$\ell$ the position of the cusp $r_{\rm {cusp}}$ is always smaller than
the marginally stable orbit $r_{\rm {ms}}$, when the angular momentum is
not constant, it may also happen that $r_{\rm {cusp}}>r_{\rm {ms}}$,
depending on the choice of the parameters ${\cal S}$ and $q$ in 
Eq.~(\ref{power_law}).

Among the large family of initial configurations that are in principle
possible, we only consider those that possess a cusp, a centre, and with
a closed equipotential through the cusp. This only happens when $0\leq
q<0.5$ and when $|{\cal S}_{\rm {ms}}|<|{\cal S}|<|{\cal S}_{\rm {mb}}|$, 
where ${\cal S}_{\rm {ms}}$ and ${\cal S}_{\rm {mb}}$ are two limiting 
values defined in \citet{daigne:04} (the subindex `mb' refers to the 
marginally bound orbit). Also note that relativistic tori with $\ell$ 
following a power-law generally have larger sizes than constant angular 
momentum tori, a fact which can produce initial models with central 
densities one or two orders of magnitude smaller.

\section{Mathematical framework}
\label{III}

	The results we report in this paper are based on two
complementary approaches involving both nonlinear hydrodynamic
simulations and the solution of the eigenvalue problem within a
perturbative linear analysis. We recall in what follows the basic
aspects related to both of them.

\subsection{Solution of the nonlinear hydrodynamics equations}

	The nonlinear, axisymmetric, general relativistic code used 
in the simulations we report is an
extended version of the one presented by \citet{zanotti:03} which
incorporates the effects of a rotating background spacetime. In this
code the hydrodynamic equations are implemented as a first-order,
flux-conservative system according to the formulation developed by
\citet{banyuls:97}. The explicit form of this system of equations for the
Kerr metric can be found in~\citet{font:02}. The equations are solved
using a high-resolution shock-capturing scheme based on an approximate
Riemann solver (either HLLE or Marquina's). Second-order accuracy in
both space and time is achieved by adopting a piecewise-linear cell
reconstruction procedure and a second order, conservative Runge-Kutta
scheme, respectively.

	The computational grid consists of $N_r \times N_{\theta}$
zones in the radial and angular directions, respectively. The innermost
zone of the radial grid is placed at $r_{\rm {min}}=r_{\rm {horizon}}+
\epsilon$, while the outermost radial grid point is at a distance about
$30\%$ larger than the outer radius of the torus, $r_{\rm {out}}$. While
$\epsilon$ depends on the particular model considered, it is typically
$0.1 r_g$ or smaller. The radial grid is built by joining smoothly
two patches obtained using two different algorithms. The first part
extends from $r_{\rm {min}}$ to the outer radius of the torus. It is
logarithmically spaced and has the maximum radial resolution at the 
innermost grid zone, $\Delta r=1\times 10^{-3}$. The spacing of the 
second part of the radial grid is uniform and it extends up to 
$r_{\rm {max}}$. The total number of radial grid points $N_r$ is selected 
so as to have a radial resolution at the outer boundary $\Delta r\simeq
0.2$. Typically, we use $N_r\simeq 250$ for the more compact tori
(radial length ${\rm L}\leq 15$), while $N_r\simeq 350$ for the more
extended models. The angular grid, on the other hand, which consists of
$N_{\theta}=84$ zones in all of the simulations, covers the domain from
$0$ to $\pi$.

	As customary for finite difference hydrodynamical codes which
cannot handle vacuum regions, a low density ``atmosphere'' is introduced
in those parts of the numerical domain not occupied by the torus. In all
our simulations we have chosen a special version of the spherically
accreting solution described by \citet{michel:72}, suitably modified to
account for the rotation of the black hole (see Appendix~A for a detailed
description). Since this atmosphere is evolved as the rest of the fluid,
one should simply take care that its dynamics does not interfere with
that of the object being studied. We have verified that this is indeed
the case if the maximum density of the atmosphere is $5-6$ orders of
magnitude smaller than the central density of the torus.

	As mentioned in the Introduction we are not interested here in
the study of the runaway instability but rather we focus on the
oscillation properties of thick discs filling their outermost closed
equipotential surface (we also refer to these discs as ``marginally
stable''). Hence, we assume that the background spacetime is simply the
one provided by the Kerr metric and that it does not change during the
evolution; this prevents the development of the instability
\citep{font:02,zanotti:03} by construction. This is a reasonable 
approximation in the present context for two different
reasons. The first one is that most of the tori we evolve have rest
masses $M_t$ which are much smaller than that of the black hole
(i.e. $M_t/M=0.1$). Hence, we can neglect their contribution to the
overall gravitational field when compared to that created by the black hole. 
The second reason is that a small power-law index $q$ in the angular
momentum distribution is enough to reduce significantly the amount of
rest-mass accreted onto the black hole so that, effectively, the mass and
spin of the black hole do not change significantly over the timescale of
our simulations. This was shown numerically by \citet{font2:02}, \citet{daigne:04} 
and explained in the perturbative analysis by \cite{ryz:03}. To validate 
this approximation we monitor the rest-mass accreted by the black hole 
according to the formula
\begin{equation}
\label{fl1}
\dot{m}(r_{\rm min})\equiv 
	- 2\pi\int_0^{\pi} \sqrt{-g}
	\rho W v^r d\theta 
	\Big\vert_{r_{\rm min}}\ ,
\end{equation}
where $g$ is the determinant of the metric and where $r_{\rm min}$ is the
radius of the innermost radial cell. Similarly, we monitor the angular momentum 
flux across the horizon as
\begin{equation}
\label{fl2}
\dot{J}(r_{\rm min})\equiv 
	- 2\pi \int_0^{\pi} \ell \sqrt{-g}
	\rho W v^r d\theta 
	\Big\vert_{r_{\rm min}}\ .
\end{equation}
Note that $W$ and $v^r$  in Eqs. (\ref{fl1}) and (\ref{fl2}) are the
Lorentz factor and the coordinate radial velocity, respectively,
as measured by the Zero Angular Momentum Observer.
In all of the simulations performed here the global relative changes in
the black hole mass and spin resulting from accretion are $\Delta M/M
\sim \Delta J/J \lesssim 10^{-4}$. Hence, the assumption of a fixed
spacetime is accurate and justified. 
\begin{table*}
\begin{center}
\caption{Fundamental properties of the initial models.  From left to
right the columns report the name of the model, the black hole spin
parameter $a$, the torus-to-hole mass ratio $M_t/M$, the power-law index $q$
of the angular momentum distribution, the parameter $\lambda$ fixing the
value of the constant ${\cal S}$, the adiabatic index $\gamma$, the polytropic
constant $\kappa$, the inner and the outer radius of the tours, $r_{\rm
in}$ and $r_{\rm out}$, the equatorial size of the torus $L$, the orbital
period at the point of maximum rest-mass density $t_{\rm orb}$, the
maximum rest-mass density, and the average
rest-mass density. For all models considered the mass of the black hole
is $M=2.5M_{\odot}$.} 
\label{tab1}
\begin{tabular}{lc|lcc|cc|cc|c|cccc}
\hline
Model   & $a$  &$M_{\rm t}/M$ & $ q $  &  $\lambda$
& $\gamma$ &$\kappa$ & $r_{\rm in}$     & $r_{\rm out}$ & $L$
&$t_{\rm orb}$ & $\rho_{\rm max}$	& $\langle\rho\rangle$  \\
& & & & & &${\rm (cgs)}$ & & & &${\rm (ms)}$  & ${\rm (cgs)}$ &${\rm (cgs)}$ &\\

\hline
A1a & 0.0  & 0.1 & 0.0  &  $^\ast$ & 4/3 &   9.66${\times} 10^{13}$ & 4.576 &
	15.889 & 11.31 & 1.86 &1.13${\times} 10^{13}$  & 
	1.78${\times} 10^{12}$  & \\
A2a & 0.0  & 0.1 & 0.05  & 0.2 & 4/3 &  1.96${\times} 10^{13}$ & 5.228 &
	11.842 & 6.61 & 1.82 & 3.90${\times} 10^{13}$  & 
	8.32${\times} 10^{12}$  & \\
A3a & 0.0  & 0.1 & 0.1  & 0.2 & 4/3 &  1.79${\times} 10^{13}$ & 5.625 &
	13.167 & 7.54 & 2.09 & 3.02${\times} 10^{13}$  & 
	6.32${\times} 10^{12}$  & \\
A4a & 0.0  & 0.1 & 0.2  & 0.5 & 4/3 &  1.23${\times} 10^{14}$ & 6.035
	&34.767 &28.73  & 3.86 &1.53${\times}10^{12}$ &
	1.69${\times} 10^{11}$  & \\
A5a & 0.0  & 0.1 & 0.2  & 0.2 & 4/3 &   1.41${\times} 10^{13}$ & 6.824 &
	17.118 & 10.29 & 2.97 &1.56${\times} 10^{13}$  & 
	3.26${\times} 10^{12}$  & \\
\hline
C1a & 0.5  & 0.1 & 0.1  & 0.8 & 4/3 &  3.47${\times} 10^{14}$ & 3.363
	&50.367 &47.0  & 0.75 & 7.84${\times}10^{12}$ &
	2.53${\times} 10^{11}$  & \\
\hline
D1a & 0.7  & 0.1 & 0.1  & 0.3 & 4/3 &  4.60${\times} 10^{13}$ & 3.103 &
	8.880 & 5.77 & 0.99 & 6.72${\times} 10^{13}$  & 
	1.26${\times} 10^{13}$  & \\
D2a & 0.7  & 0.1 & 0.1  & 0.4 & 4/3 &  9.05${\times} 10^{13}$ & 3.004 &
	11.032 & 8.03 & 1.06 & 3.36${\times} 10^{13}$  & 
	5.16${\times} 10^{12}$  & \\
D3a & 0.7  & 0.1 & 0.2  & 0.3 & 4/3 &  3.61${\times} 10^{13}$ & 3.746 &
	12.149 & 8.40 & 1.44 & 3.19${\times} 10^{13}$  & 
	5.69${\times} 10^{12}$  & \\
D4a & 0.7  & 0.1 & 0.1  & 0.2 & 4/3 &  1.86${\times} 10^{13}$ & 3.230 &
	7.202 & 3.97 & 0.91 & 1.17${\times} 10^{14}$  & 
	3.37${\times} 10^{13}$  & \\
\hline
E1a & 0.9  & 0.1 & 0.1  & 0.7 & 4/3 &  4.12${\times} 10^{14}$ & 1.971 &
	16.535 & 14.56 & 0.71 & 1.80${\times} 10^{13}$  & 
	1.04${\times} 10^{12}$  & \\
E2a & 0.9  & 0.1 & 0.1  & 0.8 & 4/3 &  4.65${\times} 10^{14}$ & 1.938 &
	25.801 & 23.86 & 0.75 & 9.17${\times} 10^{12}$  & 
	2.74${\times} 10^{11}$  & \\
E3a & 0.9  & 0.1 & 0.2  & 0.3 & 4/3 &  3.62${\times} 10^{13}$ & 2.611 &
	8.270 & 5.66 & 0.85 & 9.38${\times} 10^{13}$  & 
	1.69${\times} 10^{13}$  & \\
E4a & 0.9  & 0.1 & 0.2  & 0.4 & 4/3 &  7.16${\times} 10^{13}$ & 2.519 &
	10.459 & 7.94 & 0.91 & 4.58${\times} 10^{13}$  & 
	6.74${\times} 10^{12}$  & \\
E5a & 0.9  & 0.1 & 0.2  & 0.5 & 4/3 &  1.26${\times} 10^{14}$ & 2.443 &
	13.500 & 11.05 & 0.98 & 2.37${\times} 10^{13}$  & 
	2.69${\times} 10^{12}$  & \\
\hline
A5b & 0.0  & 0.1 & 0.2  & 0.2 & 5/3 &   1.28${\times} 10^{9}$ & 6.824 &
	17.118 & 10.29 & 2.97 &9.25${\times} 10^{12}$  & 
	3.22${\times} 10^{12}$  & \\
A5c & 0.0  & 0.1 & 0.2  & 0.2 & 2   &   9.82${\times} 10^{4}$ & 6.824 &
	17.118 & 10.29 & 2.97 &7.18${\times} 10^{12}$  & 
	3.21${\times} 10^{12}$  & \\
\end{tabular}
\begin{flushleft}
$^\ast$Note that for model A1a the constant $\lambda$ is not defined 
and $\ell=3.8$.
\end{flushleft}
\end{center}
\end{table*}

\subsection{Solution of the eigenvalue problem}
\label{seigen}

	In addition to the hydrodynamic
simulations we also perform a linear perturbative analysis of the
axisymmetric modes of oscillation of relativistic tori in a Kerr
spacetime. This is done following the procedure outlined in
\citet{ryz:03} for the Schwarzschild spacetime and recently extended to
the Kerr case by \citet{montero:04}. The method, whose details can be
found in the aforementioned references, consists in solving the perturbed
relativistic continuity and Euler equations obtained after introducing
perturbations with a harmonic time dependence in the velocity and the
pressure. The system of perturbed hydrodynamical equations is then cast
into a set of coupled ordinary differential equations and solved as an
eigenvalue problem, where the perturbed quantities are treated as
eigenfunctions and where the eigenvalues provide the eigenfrequencies of
the system. A similar procedure but for thin discs has been described
by~\citet{rodriguez:02}.

	Two approximations are adopted in the perturbative analysis. First,
we neglect the perturbations of the background spacetime, something which 
is usually referred to as the Cowling approximation \citep{cow:41}. This 
choice is consistent with that of neglecting the contribution of mass and 
angular momentum fluxes to the mass and spin of the black hole in the nonlinear 
hydrodynamic simulations. The second approximation is that of adopting a vertically
integrated description of the tori. While this approach simplifies the
numerical treatment considerably (the angular dependence is removed and
only the radial one remains) and allows for a satisfactory calculation of
the eigenfunctions and eigenfrequencies for the lowest-order modes, it
may not be sufficiently accurate to capture the properties of the higher-order 
modes, that intrinsically have a larger number of nodes in the radial and polar 
directions. This approximation produces some discrepancies between the two 
complementary approaches we use, for instance, when exciting selectively
higher-order modes, or when comparing the eigenfrequencies of modes
larger than the first overtone. A more detailed discussion of this issue is
presented in Section \ref{comparison} .

\begin{figure}
\centerline{
\psfig{file=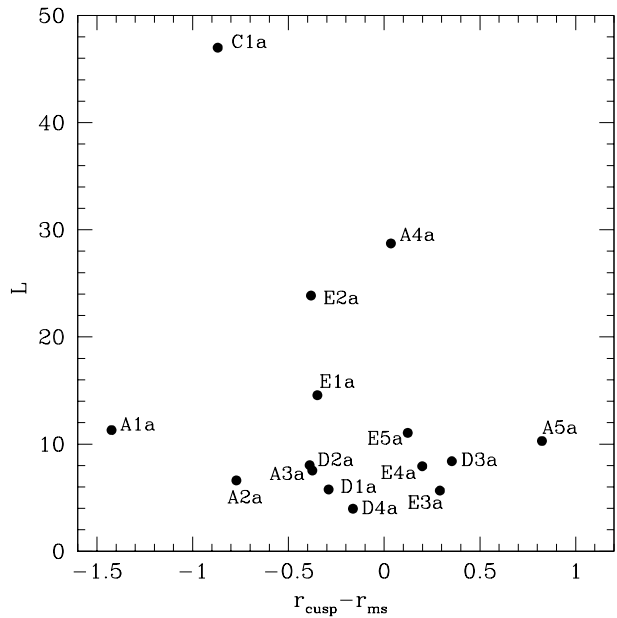,angle=0,width=8.5cm}
        }
\caption{
\label{fig1}
Distribution of the whole set of equilibrium models in the plane 
$(r_{\rm cusp}-r_{\rm ms},L)$. Notice that models A1b and A1c (not
plotted) share the same location of model A1a.}
\end{figure}
%

%
\section{Initial models}
\label{IV}

	The initial models for the numerical
simulations are a sequence of relativistic tori which fill their
outermost closed equipotential surface and thus their inner radius
coincide with the position of the cusp, i.e. $r_{\rm in}=r_{\rm cusp}$.
Table~\ref{tab1} reports the main properties of the unperturbed models.
All of the models but two (i.e. A5b and A5c) are built with an adiabatic 
index $\gamma=4/3$. More specifically, models A5b and
A5c have $\gamma=5/3$ and $\gamma=2$, respectively. 

	The value of the constant ${\cal S}$ in the power-law distribution,
Eq.~(\ref{power_law}), spans the allowed range between ${\cal S}_{\rm ms}$ and
${\cal S}_{\rm mb}$, and is selected via the additional parameter $\lambda$ as
${\cal S} = \min({\cal S}_{\rm ms},{\cal S}_{\rm mb}) + \lambda
\vert {\cal S}_{\rm mb}-{\cal S}_{\rm ms} \vert$. 
The resulting value of ${\cal S}$, together with the power-law index
$q$, are the most important factors in determining the geometrical
properties of the torus, in particular for the location of the Keplerian
points (i.e. $r_{\rm cusp}$ and $r_{\rm max}$, not reported in
Table~\ref{tab1}) and for the radial size of the disc $L \equiv r_{\rm
out} - r_{\rm in}$.

	Figure~\ref{fig1} provides a graphical view of the distribution
of the initial models in the plane $(r_{\rm cusp}-r_{\rm  ms},L)$. 
We recall that the position of the marginally stable orbit, $ r_{\rm  ms}$, 
is a decreasing function of the black hole spin parameter $a$, while by 
increasing the power law index $q$ the inner radius of the tori gets
larger. As a result it is a natural consequence of nonconstant angular 
momentum tori in the Kerr metric to have $r_{\rm cusp}-r_{\rm  ms}$ usually
larger than zero, making accretion more difficult than in the case of constant 
angular momentum tori. Only for small values of $a$ and $q$ can the torus 
penetrate deeper in the potential well, pushing the cusp below the marginally
stable orbit.

	It is worth underlining that, once perturbed, the numerical
evolution of the stationary models reported in Table~\ref{tab1} can indeed provide
an insight into the dynamics of astrophysical thick discs around
stellar-mass black holes. All of the astrophysical scenarios where
relativistic tori are expected to form (such as the merger of two neutron
stars, or of a black hole and a neutron star, or in the core collapse of
a massive star) clearly suggest that the newly formed torus
will be far from equilibrium and may indeed represent only a transient
phase to a process which will ultimately destroy the torus.

	For this reason we consider three different types of
initial perturbations, aimed at reproducing some of the physical
conditions that might occur in realistic astrophysical scenarios. The
first one has already been described and motivated in paper I and
is based on the inclusion of a radial velocity (we recall that in equilibrium
all the velocity components but the azimuthal one are zero). This
perturbation is parametrized in terms of a dimensionless
coefficient $\eta$ of the spherically symmetric accretion flow onto a
Schwarzschild spacetime~\citep{michel:72},
i.e. $v_r=\eta(v_r)_{\rm Michel}$.
 The second type of
perturbation intends to affect the rest-mass density only, either in a
global way (e.g. by rescaling the stationary profile), or in a local
random way (e.g. by introducing small variations randomly distributed
within the torus). Finally, the third type of perturbation is closely
related to the perturbative analysis discussed in Section~\ref{seigen}
and uses the eigenfunctions of the $p$-modes obtained in the linear
analysis as initial data for the perturbations in the density and
velocity. This procedure is particularly effective when the lowest-order
modes are used.

\section{Numerical results}
\label{V}

%
\begin{figure}
\centerline{
\psfig{file=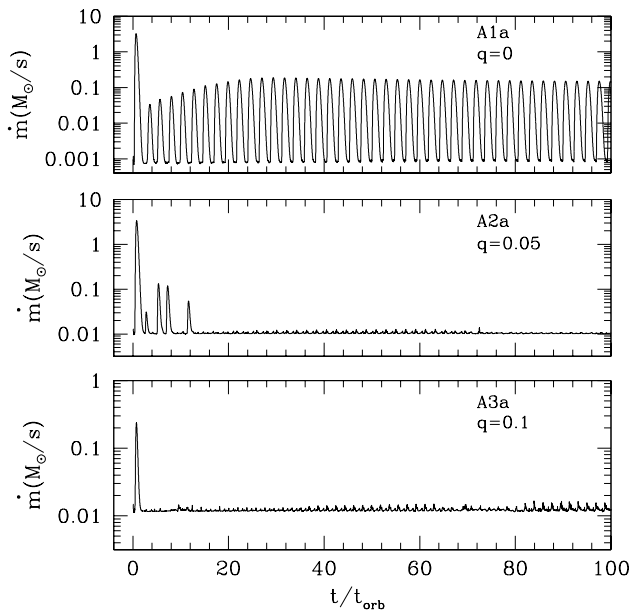,angle=0,width=8.5cm}
        }
\caption{
\label{fig63}
Time evolution of the mass accretion rate for representative
tori orbiting around a Schwarzschild  black hole. The
figure shows the stabilizing effect of increasing the
power law index from $q=0$ (model A1a) to $q=0.1$ (model
A3a), which reflects in progressively smaller accretion
rates. In all of the models the perturbation factor $\eta=0.08$.}
\end{figure}

	The disc-to-hole mass ratio considered in our 
initial models ($M_{\mathrm t}/M=0.1$) as well as the power-law index of the
angular momentum distribution chosen, make these objects stable with
respect to the runaway instability. Indeed, as reported recently by
\cite{daigne:04} for discs with $M_{\mathrm t}/M=0.1$, the critical
power-law index separating stable and unstable models appears to be
$q_{\mathrm{cr}}\sim 0.05-0.06$ for mass accretion rates of about a few
times $M_{\odot}{\rm s}^{-1}$. Hence, when perturbed, these models will
only respond with oscillations that are harmonic as long as the amplitude
of the perturbation is sufficiently small (see paper I for a discussion of 
the transition to nonlinear oscillations). As a result, with the exception 
of model A1a, which has a power law index $q=0$,
we expect the models selected in Table~\ref{tab1} to be particularly
suitable for studying their response to perturbations and, in particular,
to compute the associated mode eigenfrequencies (Section~\ref{Va}) and 
gravitational wave emission (Section~\ref{Vb}).

\subsection{Oscillation properties}
\label{Va}

\subsubsection{Dynamics of representative models}

	Figure~\ref{fig63} shows the time evolution of the accretion
rate of the first three models of Table~\ref{tab1}, namely models A1a, 
A2a, A3a, all of them referring to 
a Schwarzschild black hole. The time evolution displayed in Fig.~\ref{fig63} 
corresponds to models having an initial perturbation in the radial velocity 
with an amplitude $\eta=0.08$. The final time plotted in the figure corresponds 
to 100 orbital periods, which are measured with respect to the orbital period 
of the rest-mass density maximum in the unperturbed torus.

From model A1a to model A3a the power law index changes from $0$ to $0.1$, thus 
providing a progressive deviation from a constant angular momentum distribution 
toward a steeper power law distribution. As it is obvious from the figure 
when the power law index $q$ increases while maintaining $\lambda$ constant, 
the mass accretion rate decreases correspondingly. Already at $q=0.05$ (model 
A2a) the behaviour of $\dot{m}$ is quite different from that of a constant 
angular momentum model, and at $q=0.1$ the behaviour of the mass flux is entirely  
dominated by the behaviour of the accreting atmosphere.

We have analyzed the time evolution of the accretion rate for the rest of models 
reported in Table~\ref{tab1}, finding that apart from an early intense burst of
accretion due to the response of the system to the
initial perturbation (this is  evident e.g.~in model
A3a in Fig~\ref{fig63}), none of them shows a
significant variation of the position of the inner radius
when compared to the  initial one, thus having an essentially
suppressed mass flux. This behaviour does not
depend in a substantial manner on whether the inner
radius is smaller or larger  than the marginally stable
orbit. This is the case, for instance, of model C1a which
although has  $r_{\rm cusp}-r_{\rm ms}\sim -1.0$ is very stable and barely
accretes on to the black hole, with $\dot m \sim 10^{-6}
M_{\odot}{\rm s}^{-1}$. 
On the contrary, this behaviour strongly depends on the value of the power
law index $q$. This reflects the fact that tori with
nonconstant distributions of  specific angular momentum
have oscillations which are significantly damped at the
inner edge \citep{ryz:03}. 
We also recall that the amplitude of the
accretion rates reported in  Fig.~\ref{fig63} 
depends on the perturbation factor $\eta$, and that, as
shown in paper I, this dependence is linear for 
$\eta \lesssim 0.04$ (cf. Fig.~9 of paper I). 

\begin{figure*}
\begin{center}
\includegraphics[width=8.2cm,angle=0]{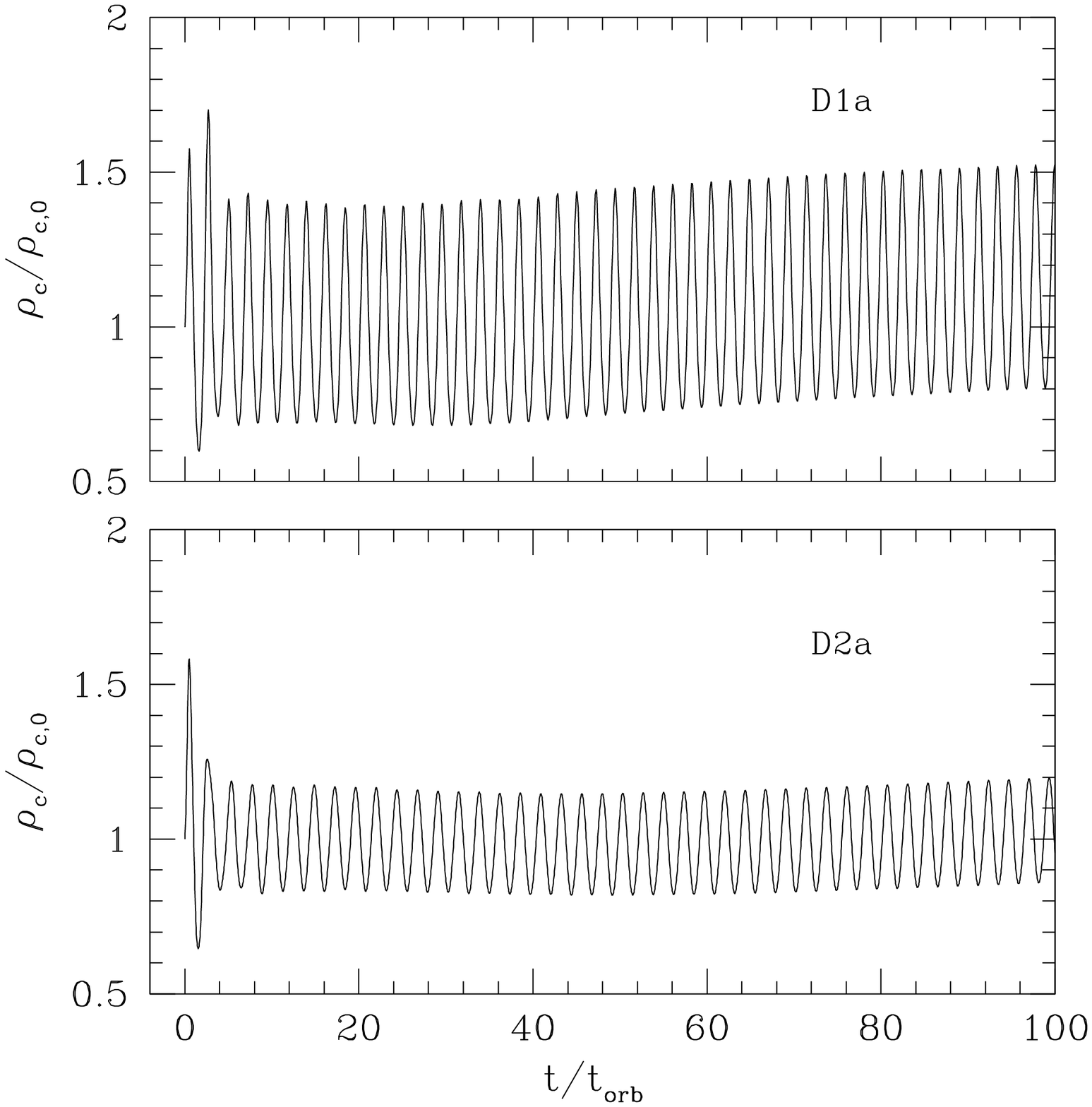}
\hspace{0.8truecm}
\includegraphics[width=8.2cm,angle=0]{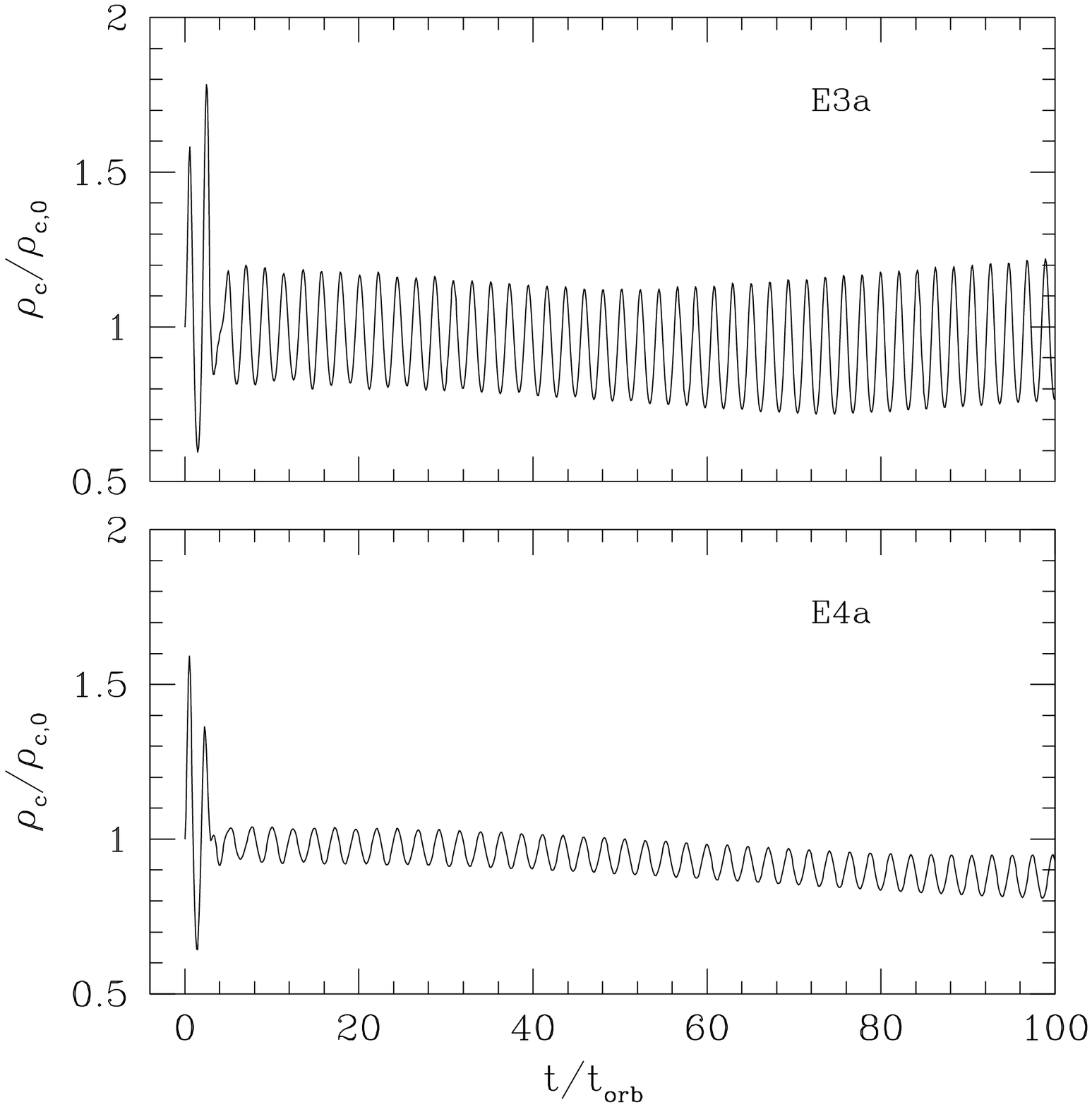}
\caption{Time evolution of the central rest-mass density normalized to
its initial value for models D1a, D2a
(black hole spin $a=0.7$) and E3a, E4a
(black hole spin $a=0.9$). The perturbation factor
$\eta=0.08$ for models of both panels.
}
\label{fig91-92}
\end{center}
\end{figure*}

Despite a distinctive quasi-periodic behaviour is quite apparent from 
Fig.~\ref{fig63}, we note that for most models the
evolution of the accretion  rate reflects mainly the
response of the atmosphere to the perturbations
propagating from the torus. 
For this reason the periodic character of the 
dynamics of the discs becomes more evident in the time evolution of the maximum
rest-mass density rather than in the mass flux.
This is reported in
Fig.~\ref{fig91-92} for four representative  models, two of them
corresponding to a Kerr black hole of spin rate $a=0.7$
(D1a, D2a), and the other two corresponding to a Kerr
black hole of spin rate $a=0.9$ (E3a, E4a).
After the tori relax from the initial perturbation at
$t/t_{\rm orb} \simeq 2$, they start oscillating at regular, quasi-periodic
intervals. This is a feature which discs with power-law distributions of 
angular momentum share with constant angular momentum discs as those 
investigated in paper I. The harmonic variation of the
hydrodynamical quantities  is analyzed in more detail next.

\subsubsection{Comparisons with the perturbative analysis}
\label{comparison}

	Additional information on the quasi-periodic behaviour of the
hydrodynamics variables discussed in the preceding section can be
extracted through a Fourier analysis of the corresponding time
evolutions. For this purpose we have calculated the Fourier transforms of
the time evolution of the $\rm {L_2}$ norm of the rest-mass density for
all models, defined as $\vert\vert\rho\vert\vert^2 \equiv
\sum_{i=1}^{N_r} \sum_{j=1}^{N_\theta}\rho^2_{ij}$. As this is a global
quantity it is particularly useful for comparisons with the results of
the eigenvalue problem in the linear perturbative analysis. Furthermore,
we can also compare our findings with those reported in paper I for the
case of constant angular momentum discs.

\begin{figure}
\centerline{
\psfig{file=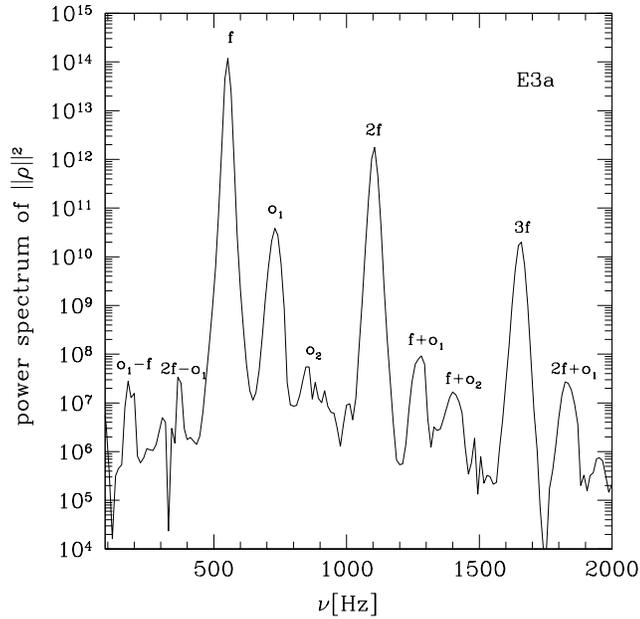,angle=0,width=8.5cm}
        }
\caption{
Power spectrum of the ${\rm L_2}$ norm of the rest-mass
density for model E3a. The units in the vertical axis are
arbitrary and the power spectrum was obtained using a Hanning filter.}
\label{fig6}
\end{figure}

	A first result that emerges from the Fourier analysis is that the
overall dynamics of nonconstant angular momentum discs is more complex
than in the case of constant distributions of the angular momentum. This
is reflected in particular in the fact that the power spectra of the
$\rm {L_2}$ norm of the rest-mass density show, in general, a richer
structure. This can be seen in Fig.~\ref{fig6}, where we plot the
power spectra of the $\rm {L_2}$ norm of $\rho$, obtained
with a Hanning filter \citep{press:96} for one representative model,
namely E3a.  

Although not all of the computed spectra show the
features of Fig.~\ref{fig6} with the same clarity,
all of them present the same essential characteristics,
namely a fundamental
mode $f$ and a series of overtones. 
Among them we can recognize the two first overtones
predicted by the linear analysis and indicated as 
$o_1$ and $o_2$. In addition to these,
further modes appear as linear combinations of $f$, $o_1$
and $o_2$, the most prominent of which  is the one at $2f$.
Since the ratio $o_1/f$ can be computed with fairly good
accuracy from the spectra, we compare it with the ratio between the two
first eigenfrequencies computed with the linear perturbative analysis. 
This comparison is reported in Table~\ref{tab2}. 
Note that, for those
models for which it was possible to identify the mode $o_1$
unambiguously, the agreement with the
prediction of the linear perturbative approach is very good, with
differences of $\sim 5\%$ at most.
\begin{table}
\begin{center}
\caption{
Frequency ratio $o_1/f$ as extracted from the power
spectra of the ${\rm L_2}$ norm of the rest mass density
 (second column) and as computed from the solution of the
eigenvalue problem (third column).}
\label{tab2}
\begin{tabular}{lcc}\hline
Model   & ${\rm (o_1/f)_{num}}$  &${\rm (o_1/f)_{linear}}$  \\
 \hline 				\\
A1a & 1.45  &  1.47 \\
A2a & 1.46  &  1.41 \\
A4a & 1.36  &  1.36 \\
A5a & 1.31  &  1.26 \\
D1a & 1.37  &  1.32 \\
D2a & 1.40  &  1.36 \\
D3a & 1.35  &  1.28 \\
D4a & 1.35  &  1.28 \\
E1a & 1.33  &  1.40 \\
E3a & 1.31  &  1.25 \\
E4a & 1.36  &  1.29 \\
E5a & 1.38  &  1.32 \\
\hline 
\end{tabular}
\end{center} 
\end{table}
\begin{figure}
\centerline{
\psfig{file=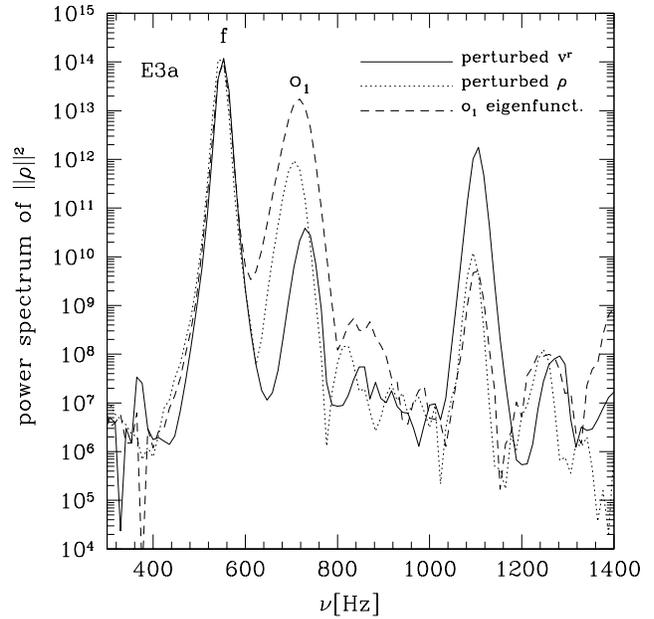,angle=0,width=8.5cm}
        }
\caption{
\label{fig5}
Power spectra of the ${\rm L_2}$ norm of the rest-mass density for model 
E3a. The different lines refer to different initial perturbations: the 
solid line corresponds to an initial perturbation given by a parametrized 
spherically symmetric radial velocity, the dotted line to a global initial
perturbation in the density, and the dashed line to an initial perturbation 
given by the eigenfunction of mode $o_1$. Vertical units are arbitrary.
The values have been rescaled in order to match the power of the fundamental 
frequency.
}
\end{figure}
A similar comparison can also be done for the mode $o_2$,
although its identification from the
computed spectra is much more difficult. In the two cases
when this was possible, namely  models E3a and
D4a, the ratio $o_2/f$ was shown to differ from
the one predicted by the linear analysis by less than $7\%$.

It is worth emphasizing that	the excitation of a
particular overtone depends sensibly  on the choice of
the initial perturbation. In particular, 
we can excite {\it
selectively} a specific mode by perturbing a given equilibrium model
through the use of the vertically integrated eigenfunctions for the
velocity and rest-mass density that have been calculated for that model
through the perturbative analysis. Exploiting this possibility we have
investigated in detail the dynamics of model E3a, considering
initial data which consist either of generic perturbations in the
radial velocity (with $\eta=0.08$) or in the rest-mass density (or in
both), or of specific perturbations obeying the eigenfunctions for the
velocity and the rest-mass density for the $o_1$ overtone (the
fundamental mode also would be excited in this case). The results of this
investigation are summarized in Fig.~\ref{fig5}, which shows the power
spectra of the $\rm {L_2}$ norm of $\rho$ for the three different initial
perturbations. As it is clear from this figure the two generic initial
perturbations, represented by a global perturbation in the radial
velocity (solid line) and by a global perturbation in the rest-mass
density (dotted line), are less efficient in exciting the mode $o_1$ 
than when the mode eigenfunction is used (dashed line). In the
latter case the power channeled in that mode increases by a factor $\sim
20$ with respect to initial perturbations in the global velocity and by a
factor $\sim 5$ with respect to the initial perturbations in the global
density.

	We note that exciting overtones of the
fundamental mode above $o_1$ is increasingly
difficult. The tests that we have performed using the eigenfunctions of
the mode $o_2$, for instance, could not provide a clear signature
of a selective excitation, in contrast with the mode $o_1$. We
believe this is probably due to the approximation made in the linear
perturbative approach in treating thick discs as vertically integrated
objects. While we expect this approximation to be a satisfactory one for
the lower-order modes, it becomes gradually less accurate as the mode
number increases and the small-scale features of the eigenfunctions
become more important.

	We have not yet commented on the presence
of the modes as linear combinations of $f$,
$o_1$ and $o_2$ shown in Fig.~\ref{fig6}, 
 and in particular on the
overtone appearing at twice the fundamental frequency,
which is, after the fundamental mode, where most of the
power is concentrated.
A plausible interpretation of the peak at $2f$, and of the
other linear combinations shown in the spectrum of Fig.~\ref{fig6},
is that they are the result of
a nonlinear coupling effect. It is, in fact, a general property of
nonlinear systems in the limit of small oscillations
that of showing nonharmonic oscillations, i.e. linear
combinations of their normal modes of oscillations. Thus, if the system
has eigenfrequencies $\omega_i$, the nonlinearity of the equations will
also produce modes at frequencies $\omega_i \pm \omega_j$ (cf. Landau \&
Lifschitz 1976, \S 28), with amplitudes  which are proportional to the product 
of the amplitudes of the combining frequencies. 
It should be noted that this nonlinear coupling is particular evident in
the present calculations which have been performed with an initial
perturbation amplitude $\eta=0.08$, at which the system's response is
already nonlinear~\footnote{The maximum perturbation amplitude triggering
linear perturbations has been estimated to be $\eta \sim 0.04$ in paper I
(cf. Fig.~9 of that paper).}.

	A consequence of this nonlinear coupling among modes is that, 
if $A_{f}$ and $A_{2f}$ denote the power spectra
amplitudes of a generic quantity $A$
(e.g. the rest-mass density)  at the frequencies 
$f$ and $2f$, respectively, we then expect
that $A_f \gg A_{2f}$ and the two amplitudes 
to scale like $A_{2f}\propto A_f^2$. We have tested that this is indeed
the case by considering one representative model, D2a. For this model we
have computed from the power spectrum a sequence of
pairs $(\log A_f,\log A_{2f})$ for different values of
the perturbation parameter $\eta$, verifying that to a
good approximation they settle along a straight line of
slope $2$ in the corresponding plane. 
\begin{figure}
\centerline{\psfig{file=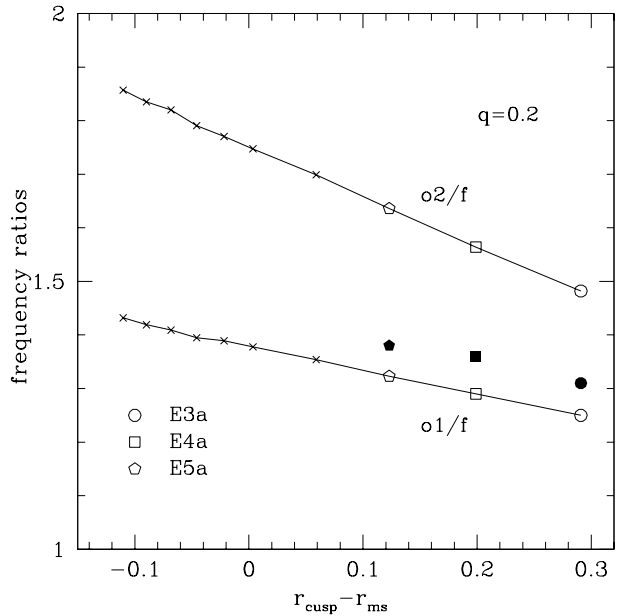,angle=0,width=8.5cm}}
\caption{
\label{fig2}
Ratio of overtones as computed through the linear analysis. The ratios 
$o1/f$ and $o2/f$ approach $3/2$ and $2$, respectively, 
as the cusp of the torus penetrates deeper in the potential well. 
Three of the models considered here (geometric symbols) belong to the
sample of Table~\ref{tab1}. For comparison, the filled
geometric symbols report the values obtained from the
numerical simulations.
}
\end{figure}

In~\citet{ryz:03} 
it was shown through linear analysis that $p$-modes in a thick disc with 
a constant distribution of angular momentum have frequencies which stay 
in the harmonic sequence $2:3:4\ldots$. 
Since this same sequence was also found 
in the nonlinear numerical simulations reported in paper I it was thus 
natural to interpret them as an evidence of the excitation of $p$-modes. 
In particular, the mode at twice the fundamental frequency was identified as 
the mode $o_2$, for which $o_2/ f=2$. 
However, it is now clear that in the case of constant angular momentum
discs the
peak at $2f$ receives contributions both from the mode $o_2$
and from the nonlinear coupling $f+f$. 
As a result,
part (if not most) of the power at 
the peak at twice the fundamental frequency in Fig.~7 of paper I 
could be the result  of the  nonlinear coupling  discussed here.

More recently, \citet{montero:04} have investigated  
$p$-modes in thick discs with more 
generic angular momentum distributions and found that in
this case
deviations from the simple sequence $2:3:4\ldots$ can be large.
We show  in Fig.~\ref{fig2} the frequency ratios
$o_1/f$ and $o_2/f$ for a number of models of our sample
as a function of the  penetration of the inner radius of
the discs in the black hole potential well.  
All of the values in Fig.~\ref{fig2} 
have been computed  with the linear code and some of the
unperturbed models coincide with those evolved with the
nonlinear code. These are E3a, E4a, E5a and are indicated
with the geometric  symbols. 
Note that both $o_1/f$ and $o_2/f$ differ
from $3/2$ and $2$, values which they approach only in
the limit of a torus penetrating deeply in the potential
well. 
Furthermore, while the
spectra obtained from the numerical simulations show a clear peak at the
position predicted by the linear analysis for the mode $o_1$, it is
difficult to find a peak at the position predicted by the perturbative
code for the $o_2$ overtone.

	In summary, the comparison between the results of the nonlinear
numerical simulations and those coming from the linear perturbation
analysis indicate that the fundamental mode $f$ and its first overtone
in the numerical simulations do represent the first two  $p$-modes of the
system and that these are in ratio $o_1/f$ close to $3/2$, with
deviations that can be as large as $\sim 15\%$. Additional $p$-modes
are also probably excited in the simulations, namely the
mode $o_2$ as shown in Fig.~\ref{fig6}
for model E3a, but the corresponding power
is in general too small to be visible in the 
spectra of the rest of the models.
On the other hand, the computed spectra
show overtones at integer multiples of the fundamental frequency (most
notably at $2f$), plus additional modes which are linear
combinations of $f$ and $o_1$, 
and that are all the result of nonlinear coupling effects. 

The present results also shed some additional light
on the physical mechanism proposed by \citet{rymz:03} to
explain HFQPOs in black hole candidates. The main idea is
that a particular class of observations showing
frequencies in the ratio $2:3$ and $1:2$ \citep{remi:02,abramowicz:03}
could be
explained in terms of the excitation of the $p$-modes of
a thick accretion disc orbiting around a black
hole. According to the numerical simulations presented
here, in the linear regime the $o_2$
$p$-mode will be little
excited, since very 
little power is in general channeled in this
mode. However, a mode at $2f$ will be present and
strongly excited in the nonlinear regime.
As a result, 
the ratio $1:2$ is satisfied with very good precision, is
independent of the distribution of the angular momentum
and appears as soon as
the nonlinear coupling is triggered.
The ratio of the fundamental mode and of the first
overtone, on the other hand, is more sensitive on the
distribution of the angular momentum and may differ from 
$2:3$ up to $15\%$.

	We conclude this section by recalling that the fundamental modes
are closely related to the epicyclic oscillations at the location of the
maximum rest mass density of the disc and that, as shown by
\citet{ryz:03} and by \citet{montero:04}, the eigenfrequency of the
fundamental $p$-mode of the disc tends in the limit of vanishing size to
the radial epicyclic frequency at the disc rest-mass density maximum.

\subsection{Gravitational wave emission}
\label{Vb}
%

	The oscillating behaviour of the perturbed tori that we have
discussed in the previous section is responsible of significant
changes of the quadrupole moment of these sources. When the tori are
compact enough and dense enough (e.g. when they are produced via
binary neutron star mergers) the changes of the quadrupole moment
result in the emission of potentially detectable gravitational radiation
\citep{zanotti:03}. In paper I it was shown that the amount of energy
that can be radiated in this way from perturbed ``toroidal neutron stars"
can indeed be very high, with resulting signal-to-noise ($S/N$) ratios
comparable or larger than those expected from the burst signal associated
with the bounce in the collapse of the core of massive stars (but see
M\"uller et al. 2004 for calculations of the gravitational radiation
resulting from convection in the core collapse scenario). This property
is closely related to the toroidal topology of these objects, which have
their maximum rest mass density off-centered and thus have
intrinsically high quadrupole moments and significant time variation of
the latter. In the following we reexamine the issue of the gravitational
wave emission from nonconstant angular momentum tori orbiting around
Kerr black holes, and compare our findings with the results of paper I.

	The procedure for computing the gravitational wave emission is
the same presented in paper I and it is based on the use of the Newtonian
quadrupole approximation. In particular, the time evolution of the wave
amplitude $A_{20}^{\rm E2} $, which is the second time derivative of the
mass quadrupole moment, is first computed through the ``stress formula''
~\citep{finn:89} to give
\begin{eqnarray}
\label{stress}
&&A_{20}^{\rm E2} =
	k \!\int\!\! \rho \biggl[v_r v^r (3 z^2 \!-\! 1) \!+\!
	v_\theta v^\theta (2 \!-\! 3 z^2) \!-\!
	v_\phi v^\phi  \nonumber \\
	& & \qquad \qquad \qquad -6 z \sqrt{(v^r v_r)(v_\theta v^\theta)
	(1 \!-\! z^2)} \biggl. - r \frac{\partial \Phi}{\partial r}
	(3 z^2 \!-\! 1) \nonumber \\
	& & \qquad \qquad \qquad\, + 
	\, 3 z\frac{\partial \Phi}{\partial \theta}
	\sqrt{1 \!-\! z^2}\biggr] r^2 {\rm d}r {\rm d}z \ ,
\end{eqnarray}
where $ k = 16 \pi^{3/2} / \sqrt{15} $, $z\equiv \cos\theta$ and $\Phi$
is the gravitational potential, approximated to the first Post-Newtonian
order as $\Phi = (1 - g_{rr})/2$. The transverse traceless (TT) wave
amplitude is then computed as \citep{zwerger:97}
\begin{equation}
\label{htt}
h^{TT}(t) = F_+ \left(\frac{1}{8}
	  \sqrt{\frac{15}{\pi}}\right) \frac{A_{20}^{\rm E2}}{R} \ ,
\end{equation}
where $R$ is the distance to the source and where the detector's beam
pattern function, $F_+=F_+(R,\theta,\phi)$, is set to one as if optimal
conditions of detectability were met. In order to investigate the
possibility that pulsating relativistic tori can be effectively detected
by the interferometric instruments presently in operation or under
construction, we have computed the signal-to-noise ratio with respect to
LIGO I, Advanced LIGO, and VIRGO. For doing this we need an estimate of
the frequency where most of the gravitational wave emission is channeled.
This is provided by the {\it characteristic frequency} $f_c$, which
is a detector dependent quantity. If the detector has a power spectral
density $S_h(f)$, then the characteristic frequency is given by~\citep{thorne:87}
\begin{equation}
f_c \equiv \left[\int_0^\infty 
	\frac{\langle|{\tilde h}(f)|^2\rangle}{S_h(f)} 
	f df \right]\left[\int_0^\infty
	\frac{\langle|{\tilde h}(f)|^2\rangle}
	{S_h(f)} df \right]^{-1} \ ,
\end{equation}
where $\langle|{\tilde h}(f)|^2\rangle$, which is an average over
randomly distributed angles of the Fourier transforms ${\tilde h}(f)$ of
$h^{TT}(t)$, has been approximated as $\langle|{\tilde h}(f)|^2\rangle
\simeq |{\tilde h}(f)|^2$. Strictly related to the characteristic
frequency is the characteristic amplitude, that gives a typical
measurement of the gravitational wave emission $h^{TT}$ detected by a
given particular instrument and reads
\begin{equation}
\label{h_c}
h_c \equiv \left[3\int_0^\infty \frac{S_h(f_c)}{S_h(f)} 
	\langle|{\tilde h}(f)|^2\rangle f df \right]^{1/2} \ .
\end{equation}
The resulting signal-to-noise ratio is then computed as
\begin{equation}
\frac{S}{N} = \frac{h_c}{h_{\rm rms}(f_c)} \ ,
\end{equation}
where ${h_{\rm rms}(f_c)} \equiv \sqrt{f_c S_h(f_c)}$ is the {\it strain
noise} of the detector at the characteristic frequency.

\begin{figure}
\centerline{
\psfig{file=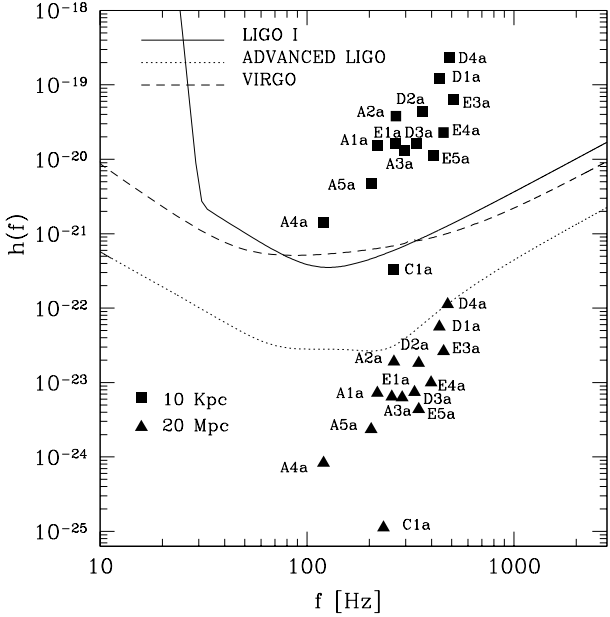,angle=0,width=8.5cm}
        }
\caption{
\label{fig30}
Characteristic wave amplitudes for the tori models of Table \ref{tab1} 
with respect to the strain noise of LIGO I and the planned sensitivity of
Advanced LIGO,  respectively. The amplitudes are computed at both a 
galactic distance of $10$Kpc and at an extragalactic distance of $20$Mpc.
The planned strain noise of VIRGO is also reported for comparison. 
}
\end{figure}

	Figure~\ref{fig30} shows the strain noise of the three detectors
considered here as a function of frequency. Note that all of the
sensitivity curves reported assume an optimally incident wave in position
and polarization (in agreement with the assumption of setting the
detector's beam pattern function to unity) and that the curve for
Advanced LIGO represents the sum of presently anticipated noise sources.
Fig.~\ref{fig30} also shows the computed characteristic amplitudes in
comparison with the strain noise of the detectors for sources located at
a distance of 10 Kpc (i.e. for galactic sources) and of 20 Mpc (i.e. for
sources within the Virgo cluster). As it is evident from the figure, in
the first case all of the models considered lie well above the
sensitivity curves of the detectors, while in the second case only one
model would be marginally detected by the Advanced LIGO
detector.  Note also that the relatively wide scattering of the models in
the figure (from the most powerful one, D4a, to the least
powerful one, C1a) are due to the differences in the average
rest-mass densities.
 
	Table~\ref{tab3} reports the characteristic frequency,
characteristic amplitude, and $S/N$ ratio of the gravitational waves
emitted by our tori models, with respect to the different detectors
considered. In particular we choose the sources at the distance of $10
{\rm Kpc}$ for LIGO I and VIRGO, while extragalactic sources at a
distance of 20 Mpc for Advanced LIGO. Also for the case of nonconstant
angular momentum discs orbiting around Kerr black holes considered in the
present paper, Table~\ref{tab3} and Fig.~\ref{fig30} confirm the results 
found in paper I for constant angular momentum discs around Schwarzschild 
black holes, namely that oscillating relativistic tori can be promising 
sources of gravitational wave emission. It is important to emphasize that 
there are firm reasons to believe that the numbers reported in 
Table~\ref{tab3} should be considered as lower limits of the amplitudes 
of the gravitational wave signals that could be effectively emitted. 
Firstly, because the torus-to-hole mass ratio could be higher than $0.1$, 
and the gravitational signal, which scales linearly with this ratio (see paper
I), would then be proportionally larger. Secondly, because the perturbations 
affecting the torus could be dramatically larger in a realistic formation 
scenario, i.e. after the core-collapse of a massive star or binary neutron 
star merger. Thirdly, because of issues having to do with the EOS of the
material of the disc. We have actually verified that by changing the
adiabatic index $\gamma$ of the polytropic EOS we use the gravitational 
radiation emitted can significantly increase. In particular, we have 
compared the wave amplitudes for models A5a, A5b, and 
A5c, which have $\gamma=4/3, 5/3$, and $2$, respectively. While 
in the transition from $\gamma=4/3$ to $\gamma=5/3$ the gravitational
signal is essentially unmodified (it increases by $\sim 3\%$), when the 
polytropic index is $\gamma=2$ (stiff EOS) the gravitational signal increases 
by $\sim 30\%$. Finally, we should also mention that Advanced LIGO has
still plenty of improvement margins achievable which are currently under
investigation [for instance, by changing the position and the transmission 
of the signal recycling mirror \citep{shoemaker:04}]. This provides an 
additional argument for considering the values reported in Table~\ref{tab3}
as lower limits of the amplitudes of the gravitational wave signal from 
relativistic thick discs.

\begin{table*}
\begin{center}
\caption{Computed estimates regarding the detection of the gravitational wave 
signals emitted by relativistic tori around Kerr black holes with power-law
distributions of the angular momentum. From left to right the table reports the 
characteristic frequency, the characteristic amplitude, and the signal-to-noise 
ratio computed for three detectors, LIGO I, VIRGO, and Advanced LIGO, assuming a galactic
distance for the first two, and an extragalactic distance for the Advanced 
LIGO detector.  $\tau_{\rm life}=100$ orbital periods.}
\label{tab3}
\begin{tabular}{ccccccccccc}
\hline
        Model & $f_c$ (Hz) & $f_c$ (Hz)& $f_c$ (Hz)& $h_c$
       & $h_c$  & $h_c$      & $S/N$     & $S/N$     &
        $S/N$ \\
       &\ LIGO I &\ VIRGO &\ ADV. LIGO    &\ LIGO I
       &\ VIRGO &\ ADV. LIGO   &\ LIGO I  &\ VIRGO &\
        ADV. LIGO &  \\
       & (10 Kpc) & (10 Kpc)  & (20 Mpc) & (10Kpc)  & (10Kpc)
       & (20 Mpc) & (10 Kpc) & (10 Kpc)  & (20 Mpc)      \\
\hline
A1a & $223$ &$236$ &$223$ & $1.5\times
10^{-20}$ & $1.5\times 10^{-20}$ & $7.1\times 10^{-24}$ &
$29.1$  & $23.5$ &  $0.26$ \\
A2a & $267$ &$274$ &$264$ & $3.8\times
10^{-20}$ & $3.8\times 10^{-20}$ & $1.9\times 10^{-23}$ &
$62.6$  & $55.2$ &  $0.59$ \\
A3a & $296$ &$308$ &$291$ & $1.3\times
10^{-20}$ & $1.4\times 10^{-20}$ & $6.3\times 10^{-24}$ &
$18.9$  & $17.7$ &  $0.17$ \\
A4a & $121$ &$117$ &$122$ & $1.4\times
10^{-21}$ & $1.7\times 10^{-21}$ & $8.3\times 10^{-25}$ &
$4.1$  & $3.2$ &  $0.03$ \\
A5a & $205$ &$207$ &$206$ & $4.6\times
10^{-21}$ & $4.7\times 10^{-21}$ & $2.3\times 10^{-24}$ &
$10.1$  & $7.5$ &  $0.09$ \\
C1a & $269$ &$388$ &$237$ & $3.3\times
10^{-22}$ & $4.2\times 10^{-22}$ & $1.1\times 10^{-25}$ &
$0.5$  & $0.5$ &  $<10^{-2}$ \\
D1a & $446$ &$454$ &$436$ & $1.2\times
10^{-19}$ & $1.2\times 10^{-19}$ &$5.6\times 10^{-23}$ & $100$
 & $122$ & $0.6$  \\
D2a & $366$ &$393$ &$349$ & $4.3\times
10^{-20}$ & $4.5\times 10^{-20}$ &$1.8\times 10^{-23}$ & $48$
 & $52$ & $0.3$  \\
D3a & $344$ &$367$ &$336$ & $1.6\times
10^{-20}$ & $1.7\times 10^{-20}$ &$7.3\times 10^{-24}$ & $19.5$
 & $20.4$ & $0.14$  \\
D4a & $498$ &$505$ &$487$ & $2.3\times
10^{-19}$ & $2.4\times 10^{-19}$ &$1.1\times 10^{-22}$ & $175$
 & $225$ & $1.0$  \\
E1a & $272$ &$341$ &$258$ &$1.6\times
10^{-20}$ &$1.9\times 10^{-20}$ &$6.4\times 10^{-24}$ & $26.4$
 & $23.8$ & $0.21$  \\
E2a & $249$ &$323$ &$236$ &$8.5\times
10^{-21}$ &$1.1\times 10^{-20}$ &$3.3\times 10^{-24}$ & $15.1$
 & $13.2$ & $0.12$  \\
E3a & $538$ &$556$ &$506$ &$6.5\times
10^{-20}$ &$6.7\times 10^{-20}$ &$3.0\times 10^{-23}$ & $44$
 & $58$ & $0.23$  \\
E4a & $454$ &$493$ &$406$ &$2.3\times
10^{-20}$ &$2.4\times 10^{-20}$ &$1.0\times 10^{-23}$ & $20$
 & $24$ & $0.12$  \\
E5a & $412$ &$491$ &$347$ &$1.1\times
10^{-20}$ &$1.3\times 10^{-20}$ &$4.3\times 10^{-24}$ & $10.9$
 & $12.3$ & $0.07$  \\
\hline
\end{tabular}
\end{center}
\end{table*}

\section{Conclusions}
\label{VI}

	We have studied the dynamics and the gravitational wave emission
from nonconstant specific angular momentum tori undergoing axisymmetric 
oscillations while orbiting around Kerr black holes. The angular
momentum distribution has been chosen to be increasing outward with the
radial distance following a power-law dependence. The self-gravity of the
discs has been neglected and the accretion of mass and angular momentum
has been assumed not to affect the underlying background metric. We have
also removed the possible role played by the runaway instability by
selecting indeces in the power-law distribution of the specific angular
momentum that are above the critical value for the instability
(cf.~\citet{daigne:04}). The results presented in this paper have thus 
extended the previous investigation of~\citet{zanotti:03} where constant 
specific angular momentum discs around Schwarzschild black holes were considered.

	A comprehensive sample of initial (marginally stable) equilibrium
tori has been built. These tori have been subsequently perturbed in
various ways in order to study their dynamical response to the
perturbations during long-term time evolutions (extending up to 100
orbital periods for each model). The time evolutions are characterized by
a distinctive quasi-periodic pattern present in all fluid quantities as
well as in the mass and angular momentum accretion rates and in the
gravitational waveforms (which have been extracted using the Newtonian
quadrupole formula). Our study has been carried out using two
complementary tools: an axisymmetric nonlinear hydrodynamics code and a
linear perturbative code for vertically integrated axisymmetric
tori. While the first one has proven useful to investigate nonlinear
effects in the axisymmetric oscillation, the second approach has been
useful in confirming and interpreting the numerical simulations.

	The main results of our study are as follows: The power-law
angular momentum distribution results in more stable tori than in the
uniform angular momentum case, yielding systematically smaller values of
the mass accretion rate as the power-law index increases. This is in 
agreement with the previous studies of \citet{ryz:03} and
of \citet{daigne:04}.
The comparison of the eigenfrequencies calculated through the perturbative 
analysis and those obtained from the power spectra of the hydrodynamical 
variables in the numerical simulations have shown a good agreement between 
the two approaches for the first two modes $f$ and $o_1$. They have been 
found close to a simple sequence of integers 2:3, but with deviations that 
become significant as the power-law index $q$ increases. Namely, for $q=0.2$ 
the ratio $f/o_1$ differs from $2/3$ by $\sim 15\%$. On the other hand, the 
second overtone $o_2$ predicted by the perturbative analysis could  be 
detected only in two cases among the several power spectra calculated from the 
hydrodynamical models. 
This is either due to the limitations of the 
vertically integrated assumption in the perturbative analysis in capturing 
higher frequency oscillations (something which is also confirmed when 
performing selective excitations of modes), or by the very small power 
confined in these modes which prevents them to become visible in the power 
spectra.

	The comparison has also shown that the peak found at $2f$ in the
power spectra should not be interpreted as a proper eigenmode of the system.
Rather, it represents the result of a nonlinear
coupling of the lower-order modes as it is common in nonlinear systems
subject to nonharmonic oscillations.

	Finally, we have reported on the detectability of the
oscillations of perturbed tori via their gravitational wave emission. In
good agreement with what we already found for constant angular momentum
discs, the chances for detection of gravitational radiation from
nonconstant angular momentum tori, when sufficiently compact and dense,
are good and within the sensitivity curves of LIGO and VIRGO for galactic
sources, but only marginal even for Advanced LIGO for extragalactic sources
located at 20 Mpc.

	Extensions of the work reported here include the incorporation of
additional physics in the models such as viscosity and magnetic fields
and will be reported in forthcoming papers.

\section*{Acknowledgments}

	It is a pleasure to thank David Shoemaker for kindly providing 
an updated sensitivity curve for the Advanced LIGO detector. We also thank 
Frederic Daigne for discussions.
O.Z. acknowledges financial support from a fellowship by the Spanish MECD
(SB2002-0128). J.A.F. acknowledges financial support from the Spanish
Ministerio de Ciencia y Tecnolog\'{\i}a (grant AYA 2001-3490-C02-01). The
computations were performed on the Beowulf Cluster for Numerical Relativity 
{\tt Albert100}, at the University of Parma and on the SGI/Altix computer 
{\tt CERCA} of the Department of Astronomy and Astrophysics of the University 
of Valencia.

\appendix
\label{appendix_A}
\section{Description of the atmosphere used in the numerical simulations}

	We model the low density atmosphere surrounding the tori in terms
of a semi-analytic solution of the relativistic stationary accretion of
a rotating fluid onto a Kerr black hole. In this sense, it represents the
extension of the relativistic spherical accretion solution onto a
Schwarzschild black hole \citep{michel:72} to account for the rotation of
both the infalling fluid and of the background spacetime.

	In practice, in addition to the continuity
equation, $\nabla_\alpha (\rho
u^\alpha)=0$, and to the energy conservation equation, $\nabla_\alpha
T^\alpha_t=0$, one also has to ensure the conservation of the angular
momentum, $\nabla_\alpha T^\alpha_\phi=0$. During its motion, a fluid with
a general four velocity $u^\alpha\equiv (u^t,u^r,0,u^\phi)$ follows 
a spiral ending at the black hole horizon along cones of
constant $\theta$. 
This motion can be derived from the following three constraints coming directly
from the conservation equations
\begin{eqnarray}
\label{m1}
\sqrt{-g}\rho u^r &=& C_1, \\
\label{m2}
\sqrt{-g}h\rho u^r u_t&=& C_2, \\
\label{m3}
\sqrt{-g}h\rho u^r u_\phi&=& C_3, 
\end{eqnarray}
where $C_1$, $C_2$ and $C_3$ can only depend on the polar angle $\theta$.
Dividing Eqs.~(\ref{m3}) and (\ref{m2}) gives the condition of constant
specific angular momentum $\ell=- u_\phi/u_t$, while the division of Eqs.~(\ref{m2}) 
and (\ref{m1}) provides the relativistic Bernoulli equation
\begin{equation}
\label{mb}
h u_t = C \ ,
\end{equation}
where $C=C_2/C_1$.
Following~\citet{michel:72}, Eq.~(\ref{mb}) is squared and
then differentiated to obtain, with the help of the
continuity equation, Eq.~(\ref{m1}), the so-called ``wind equation"
\begin{eqnarray}
\label{wind1}
&& 2\frac{du}{u}\left[-V^2 G(r,u) +
\frac{g_{rr}}{H(r,u)}u^2\right] + 
\nonumber \\
\label{wind2}
&&\frac{dr}{r}\left[-4 V^2 G(r,u) + S(r,u)\right]=0 \ ,
\end{eqnarray}
where $u$ is a shorthand notation for $u^r$, while 
\begin{eqnarray}
F(r,u) &=& 1+g_{rr} u^2 \\
H(r,u) &=& g^{tt} - 2\ell g^{t\phi} + \ell^2 g^{\phi\phi}
\\
G(r,u) &=& F(r,u)/H(r,u) \\
S(r,u) &=& r[ u^2 \partial_r g_{rr} H(r,u) - F(r,u) \\
&& (\partial_r g^{tt} - 2\ell \partial_r g^{t\phi} + \ell^2
\partial_r g^{\phi\phi})]/H(r,u)^2 \ ,
\end{eqnarray} 
and $V^2=d\ln(h\rho)/d\ln\rho - 1$ is the square of the sound speed.  
Imposing the vanishing of the terms in the square brackets of Eq.~(\ref{wind2})  
allows to determine the position of the critical point of the flow, thus
guaranteeing that the derivative $du/dr$ is always finite. Note that according 
to Eq.~(\ref{wind2}) the sound speed at the critical point is given by
\begin{equation}
\label{sound}
V^2 = \frac{g_{rr} u^2}{1+g_{rr}u^2} \ ,
\end{equation}
which is different from the velocity of the fluid $v^2=g_{ij}v^i v^j$. This
means that the critical point is not a transonic point as in the case of 
purely spherical accretion.

In the numerical implementation of this solution we further impose the 
simplifying assumption $u_\phi/u_t=0$, thus reducing considerably the
algebra. Once the density $\rho_c$ at the critical point has been given 
as a free parameter, the rest of the solution is computed as follows:
\begin{enumerate}
\item Compute $V_c$ from the equation of state. 
\item Compute the relation between $u_c$ and $r_c$ 
by equating the terms in the square
brackets of Eq.~(\ref{wind2}).
\item Compute the position of the critical point $r_c$ by
solving Eq.~(\ref{sound}).
\item Compute the constants $C_1$,$C_2$, and $C_3$ from the
known values of the solution at the critical point.
\item Solve the relativistic Bernoulli equation as an
algebraic equation in the unknown $u(r)$. Complete the
solution by computing $\rho(r)$ from the continuity
equation and all of the other thermodynamic quantities
from the equation of state.
\end{enumerate}

\bibliographystyle{mn2e}


\begin{thebibliography}{}

\bibitem[\protect\citeauthoryear{{Abramowicz}}{{Abramowicz}}{1971}]
{abramowicz:71} {Abramowicz} M.~A., 1971, Acta Astron.,
21, 81


\bibitem[\protect\citeauthoryear{{Abramowicz}, {Calvani} \&
{Nobili}}{{Abramowicz} et~al.}{1983}]{abramowicz:83} {Abramowicz} M.~A.,
{Calvani} M., {Nobili} L., 1983, Nature, 302, 597

\bibitem[\protect\citeauthoryear{{Abramowicz}, {Bulik} 
{Bursia} \& {Kluzniak}}{{Abramowicz}
    et~al.}{2003}]{abramowicz:03} {Abramowicz} M.~A.,
  {Bulik} T., {Bursa} M., {Kluzniak} W. 2003, A\&A, 404, L21-L24


\bibitem[\protect\citeauthoryear{{Abramowicz},{Klu\'zniak}} {{Abramowicz
\& Klu\'zniak}}{2004}]{ak:04}{Abramowicz, M.A., Klu\'zniak, W., 2004,
Proceedings of {\it ``X-ray Timing 2003: Rossi and Beyond''},
ed. P. Kaaret, F.K. Lamb, and J.H. Swank}


\bibitem[\protect\citeauthoryear{{Aloy}, {M{\" u}ller}, {Ib{\' a}{\~
n}ez}, {Mart\'{\i}} \& {MacFadyen}}{{Aloy} et~al.}{2000}]{aloy:00} {Aloy}
M.~A., {M{\" u}ller} E., {Ib{\' a}{\~ n}ez} J.~M.,
{Mart\'{\i}} J.~M., {MacFadyen} A., 2000, ApJ, 531, L119

\bibitem[\protect\citeauthoryear{{Aloy}, {Janka} \& {M{\" u}ller}}
{{Aloy} et~al.}{2004}]{aloy:04} {Aloy}
M.~A., Janka, H.-Th., {M{\" u}ller} E., 2004, astro-ph/0408291

\bibitem[\protect\citeauthoryear{{Balbus}}{{Balbus}}{2003}]
{balbus:03}{Balbus} S. A., Ann. Rev. Astron. Astrophy., 41, 555-597

\bibitem[\protect\citeauthoryear{{Banyuls}, {Font}, {Ib\'a\~nez},
{Mart\'{\i}} \& {Miralles}}{{Banyuls} et~al.}{1997}]{banyuls:97}
{Banyuls} F., {Font} J.~A., {Ib\'a\~nez} J.~M., {Mart\'{\i}} J.~M.,
{Miralles} J.~A., 1997, ApJ, 476, 221








\bibitem[\protect\citeauthoryear{{Cowling}}{Cowling}{1941}]{cow:41}{Cowling}
T.G.,  1941, MNRAS, 101, 367



\bibitem[\protect\citeauthoryear{{Daigne} \& {Font}}{{Daigne} \&
{Font}}{2004}]{daigne:04} {Daigne} F., {Font} J.~A., 2004,
MNRAS, 349, 841


\bibitem[\protect\citeauthoryear{{De Velliers}, {Hawley}, \&
{Krolik}}{{De Villiers} et~al.}{2003}]{devel:03} {De Velliers} J.~P.,
{Hawley} J.~F., {Krolik} J.~H., 2003, ApJ, 599, 1238






\bibitem[\protect\citeauthoryear{{Finn}}{1989}]{finn:89}
{Finn} L. S. in {\it Frontiers in Numerical Relativity},
ed. C.R.~Evans, S. L.~Finn., \& D.W.~Hobill,
Cambridge University Press, Cambridge, England, 1989








\bibitem[\protect\citeauthoryear{{Font} \& {Daigne}}{{Font} \&
{Daigne}}{2002a}] {font:02}{Font} J.~A., {Daigne} F., 2002a, MNRAS, 334,
383

\bibitem[\protect\citeauthoryear{{Font} \& {Daigne}}{{Font} \&
{Daigne}}{2002b}] {font2:02}{Font} J.~A., {Daigne} F., 2002b, ApJ, 581,
L23




\bibitem[\protect\citeauthoryear{{Gammie},{McKinney},{Toth}}{{Gammie}, {McKinney} \&
{T\'oth}}{2003}]{harm} {Gammie, C.~F., McKinney, J.~C., T\'oth, G., 2003, ApJ,
589, 444}

\bibitem[\protect\citeauthoryear{{McKinney}{Gammie}}{{McKinney} \&
{Gammie}}{2002}]{mck:02} {McKinney,J.~C., Gammie, C.~F., 2002, ApJ, 573, 728}

\bibitem[\protect\citeauthoryear{{Gammie},{Shapiro}\&{McKinney}}
{{Gammie}, {Shapiro} \& {McKinney}}{2004}]
{gammie:04} {Gammie, C.~F., Shapiro, S.~L., McKinney, J.~C., 2004, ApJ, 602, 312}

\bibitem[\protect\citeauthoryear{{Goodman}}{{Goodman}}{1986}]{goodman:86}
{Goodman} J., 1986, ApJ, 308, L46

\bibitem[\protect\citeauthoryear{{Hirose},{Krolik},{De
      Velliers},{Hawley}}{{Hirose} et~al.}{2003}]{hirose:03} {Hirose} S., 
{Krolik} J.~H., {De Velliers} J.~P., {Hawley} J.~F. 2003, astro-ph/0311500









\bibitem[\protect\citeauthoryear{{Kozlowski}, {Jaroszynski} \&
{Abramowicz}}{{Kozlowski} et~al.}{1978}]{kozlowski:78} {Kozlowski} M.,
{Jaroszynski} M., {Abramowicz} M.~A., 1978, A\&A, 63, 209


\bibitem[\protect\citeauthoryear{{Landau} \&
{Lifschitz}}{1976}]{landau:76} {Landau} L. D., \& {Lifshitz} E. M., {\it
Mechanics}, 1976, Oxford, Pergamon Press




\bibitem[\protect\citeauthoryear{{Lee}}{{Lee}}{2001}]
  {lee:01}{Lee} W.~H., 2001, MNRAS, 328, 583





\bibitem[\protect\citeauthoryear{{MacFadyen} \& {Woosley}}{{MacFadyen} \&
{Woosley}}{1999}] {macfadyen:99}{MacFadyen} A.~I., {Woosley} S.~E., 1999,
ApJ, 524, 262



\bibitem[\protect\citeauthoryear{{Masuda}, {Nishida} \&
{Eriguchi}}{{Masuda} et~al.}{1998}]{masuda:98} {Masuda}
N., {Nishida} S., {Eriguchi} Y., 1998, MNRAS, 297, 1139




\bibitem[\protect\citeauthoryear{Michel}{Michel}{1972}]{michel:72} Michel
F., 1972, Astrophys. Spa. Sci., 15, 153




\bibitem[\protect\citeauthoryear{Montero, Rezzolla \& Yoshida}{Montero
et~al.}{2004}]{montero:04} {Montero} P.J., {Rezzolla} L., {Yoshida} S'i.,
MNRAS, {in press}

\bibitem[\protect\citeauthoryear{M\"uller, Rampp, Buras, Janka \&
Shoemaker}{M\"uller et~al.}{2004}]{mueller:04} {M\"uller} E., {Rampp} M.,
{Buras} R., {Janka} H.-Th., {Shoemaker} D.H., 2004, ApJ, 603, 221





\bibitem[\protect\citeauthoryear{{Paczynski}}{{Paczynski}}{1986}]
{paczynski:86} {Paczynski} B., 1986, ApJ, 308, L43

\bibitem[\protect\citeauthoryear{{Press}, {Teukolsky},
    {Vetterling} \& {Flannery}}{Press et~al.}{1996}]{press:96}
{Press} W. H., {Teukolsky} S. A., {Vetterling} W. T. \&
    {Flannery} B. P., in {\it Numerical Recipes. The art
    of scienfitic computing}, Cambridge
University Press, 1996

\bibitem[\protect\citeauthoryear{{Remillard}, {Muno},
    {McClintock}, \& {Orosz}}{{Remillard}
    et~al.}{2002}]{remi:02} 
{Remillard} R.A., {Muno} M.P., {McClintock} J. E.,
    {Orosz} J. A., 2002, ApJ, 580, 1030

\bibitem[\protect\citeauthoryear{{Rezzolla}, {Yoshida}, {Maccarone} \&
{Zanotti}}{{Rezzolla} et~al.}{2003a}]{rymz:03} {Rezzolla} L.,
{Yoshida} S'i., {Maccarone} T. J., {Zanotti} O., 2003a MNRAS, 344, L37

\bibitem[\protect\citeauthoryear{{Rezzolla}, {Yoshida} \&
{Zanotti}}{{Rezzolla} et~al.}{2003b}]{ryz:03} {Rezzolla} L., {Yoshida}
S'i., {Zanotti} O., 2003b, MNRAS, 344, 978

\bibitem[\protect\citeauthoryear{{Rodr\'{\i}guez}, {Silbergleit} \&
{Wagoner}}{{Rodr\'{\i}guez} et~al.}{2002}]{rodriguez:02} {Rodr\'{\i}guez} M.~O.,
{Silbergleit} A.~S., {Wagoner} R.~V., 2002, ApJ, 567, 1043 


\bibitem[\protect\citeauthoryear{{Ruffert} \& {Janka}}{{Ruffert} \&
{Janka}}{2001}]{ruffert:01} {Ruffert} M., {Janka} H.-T., 2001, A\&A, 380,
544






\bibitem[\protect\citeauthoryear{{Shibata}, {Taniguchi} \&
{Ury{\=u}}}{{Shibata} et~al.}{2003}]{shibata:03}
{Shibata} M., {Taniguchi} K., {Ury{\= u}} K., 2003,
Phys. Rev. D, 68, 084020

\bibitem[\protect\citeauthoryear{Shoemaker}{Shoemaker}{2004}]{shoemaker:04} 
{Shoemaker} D. H., 2004, {\em private communication}


\bibitem[\protect\citeauthoryear{{Thorne}}{1987}]{thorne:87}
{Thorne} K. S. in {\it 300 Years of Gravitation},
edited by S. W. Hawking ans W. Israel, Cambridge
University Press, Cambridge, England, 1987


\bibitem[{{Woosley}(2001)}]{woosley:01}
{Woosley}, S.~E. 2001, in Gamma-ray Bursts in the Afterglow Era, Proceedings of
  the International workshop held in Rome, CNR headquarters, 17-20 October,
  2000, ed. E.~{Costa}, F.~{Frontera}, \& J.~{Hjorth} (Berlin Heidelberg:
  Springer), 257

\bibitem[\protect\citeauthoryear{{Yokosawa}}{{Yokosawa}}{1995}]{yokosawa:95}
{Yokosawa} M., 1995, PASJ, 47, 605

\bibitem[\protect\citeauthoryear{{Zanotti}, {Rezzolla}, \&
{Font}}{{Zanotti} et~al.}{2003}]{zanotti:03} {Zanotti} O.,
{Rezzolla} L., {Font} J~.A., 2003, MNRAS, 341, 832


\bibitem[\protect\citeauthoryear{{Zwerger} \& {M\"uller}}{{Zwerger} \&
{M\"uller}}{1997}]{zwerger:97} {Zwerger} T., {M\"uller}
E., 1997, A\&A, 320, 209

\end{thebibliography}

\label{lastpage}  
\end{document}